\documentclass[smallextended]{svjour3}       
\smartqed  
\usepackage{hyperref}
\hypersetup{colorlinks=true, urlcolor=blue, citecolor=blue, linkcolor=blue}
\usepackage{graphicx}
\usepackage{amssymb,epstopdf}

\journalname{Int. J. Theo. Phys.}
\begin{document}

\title{Dynamics of an atom cavity field system in interacting Fock space}


\author{P. K. Das \and Arpita Chatterjee*}

\authorrunning{P. K. Das \and Arpita Chatterjee} 

\institute{*{Corresponding Author}\\\\
P. K. Das\at
              Physics and Applied Mathematics Unit\\Indian
Statistical Institute\\203, B. T. Road, Kolkata 700108\\
\email{das.daspk@gmail.com}
\and
A. Chatterjee \at
              Department of Mathematics\\
              J. C. Bose University of Science and Technology\\
              YMCA, Faridabad\\
              \email{arpita.sps@gmail.com}           
                      }

\date{Received: date / Accepted: date}
\maketitle

\begin{abstract}
In this paper, we investigate one-time passing of a $V$-type three-level atom through a single-mode interacting field in a cavity. We extend the idea of elementary Jaynes-Cummings model by assuming that the field vector belongs to interacting Fock space. In the process, we arrive at a state vector which will be analyzed to study the nonclassicality of the evolved state of the system.

\end{abstract}

\section{Introduction}
\label{intro}

A manifold of nonclassical features of quantum light has received a notable attention in quantum optics \cite{girish,das5,peng1} community for a number of reasons. For example, squeezed state can be used to reduce the noise level in one of the phase-space quadrature below the quantum limit \cite{ekert}. Squeezed state is also used in continuous variable quantum cryptography \cite{hillery00}, teleportation of coherent state \cite{akira98} etc. Specifically, in the LIGO experiment, squeezed vacuum state is successfully used for the detection of the gravitational waves by reducing the noise \cite{abbott16,harry10,grote13}. Entangled states produced in down-conversion process can be employed to test fundamental aspects of quantum physics, such as non-locality \cite{zeilinger}. These states have been proved to be useful for various quantum information tasks such as quantum teleportation, dense-coding, quantum cryptography etc. \cite{kishore14,kishore17,nasir16}. Photon anti-bunching can be visualized as a state of light field in which photons prefer to travel alone in comparison to traveling in a group \cite{loudon}. Antibunched light exhibits nonclasscality \cite{malvin}. Sunlight and light used at home are in bunched state whereas some light sources (e.g. lasers) neither show any preference for traveling alone nor for travelling in a company of other photons \cite{ghatak}. Such a state of light is considered as coherent state \cite{nasirn1}.  Anti-bunching is used for characterizing single photon sources \cite{lounis} which are essential for the realization of various schemes for secure quantum communication. Not only that, by employing nonclassical light sources \cite{nasirn2} the performances of optical technology such as metrology, communication and imaging can be improved beyond the limitation of classical physics. In another aspect, the preparation of quantum entangled states through a cavity QED is a subject of intense theoretical and experimental studies. Analyzing such states evokes insight into the fundamentals of quantum mechanics. Manipulation of a light field at the single-photon level \cite{nasirn3} provides a basis for important applications in quantum information science. A desired field state can be obtained by applying two elementary operations on a single-mode field \cite{nasirn4}. Using photon addition or subtraction or their superposition, one can generate a suitable nonclassical state from any classical state which is very useful in quantum information processing. For example, both the photon-subtracted and photon-added squeezed states are suggested to improve fidelity of continuous variable teleportation \cite{lee1}. Thus manufacturing and handling nonclassical correlations in complex atomic systems united with radiation fields is one of the most challenging features of quantum information theory (QIT). In this context, we consider a three-level atomic system for studying the quantum features of the semiclassical atom-field system.

Jaynes-Cummings model helps us to understand the interaction between a single atom with a high-quality cavity \cite{jay,sten,eberly}. The interaction of an atom and a laser beam in a cavity performed close to one of the atomic resonances works as a resource of light emission with a rich set of spectral and temporal features. Temporally, the emitted light shows anti-bunching with a second order correlation. Spectrally, with the increase of laser intensity, light emitted has symmetric side lobes around the central excitation frequency. This paper is also motivated by the fact that Fock states and superposition of Fock states can be produced in cavity QED by using resonant interactions of two or three-level atoms, one at a time, with a cavity mode. The production of a two-photon state in a high-$Q$ cavity is also reported \cite{martin}. A recent paper has described a deformed atom-cavity field system constructed from the standard Jaynes-Cummings model by transforming the field operators and adding a nonlinear Kerr-like medium \cite{arpita19}. The zero space-time dimensional case of the interacting Fock space corresponds to a non-linear deformation of the usual (Boson, Fermion or q-deformed) one mode Fock spaces \cite{accar1}. The concept of quantum entanglement in the stochastic limit goes beyond the familiar notion of superposition and leads to the conclusion that under appropriate physical situations, nonlinearly interacting quantum system cannot be separated even at a kinematical level and behave as a single new quantum object satisfying new types of commutation relations and therefore new statistics \cite{accar2}. The mathematical counterpart of this qualitative statement moves towards the formulation of interacting Fock space that has been developed for describing the state space of the interacting systems. Interacting Fock space is a generalized algebraic construction used in quantum mechanics to build the quantum state space of a variable or unknown number of identical particles from a single particle Hilbert space $H$. For example, a so-called one-mode interacting Fock space is $H=\mathbb{C}$.

The nonclassical properties \cite{nasir1} of three-level atomic systems have been well studied in quantum optics for understanding quantum-coherence phenomena such as electromagnetically induced transparency (EIT) \cite{fle05}, lasing without inversion \cite{scully1}, and coherent trapping \cite{scully2}. Three-level atoms interacting with low strength driving fields, similar to EIT systems, have been used to generate entangled two-mode photon states which can be suitably manipulated to yield desired correlations \cite{sandhya}. The passage of two $V$-type three-level atoms through a cavity field have been resulted to transfer the classical cavity field into a nonclassical one \cite{arpita12}. The knowledge of nonclassical correlations carried by emitted photons in atomic systems may prove immensely useful in designing future QIT systems for communications and computations. However, the nonclassical nature of an atom-filed structure in interacting Fock space (a weighted Hilbert space with weights $\{\lambda_n\}$) is less reported, which motivated us to study the dynamics of nonclassicality parameters in interacting Fock space.

In this paper, we consider an interacting one-mode field which interacts in a cavity with the atom by letting a $V$-type atom  passing through it. After tracing out the atomic part from the generated atom-field system we obtain the field left in the cavity and explore the nonclassical properties of the field.

In the beginning, we describe the basic idea of one-mode interacting Fock space. Then we give the time-dependent state of the system containing a $V$-type three-level atom \cite{das5,peng1} which interacts with a single-mode of interacting field. In subsequent sections we show nonclassicality of the evolved state with the help of Mandel's $Q$ parameter and the squeezing properties of the radiation field. Lastly, we give a conclusion.

\section{Basic preliminaries and notations}
\label{sec:1}

As a vector space one-mode interacting Fock space $\Gamma(I\!\!\!\!C)$ \cite{luigi1,luigi2,luigi3,luigi4,das1,das2,das3,das4} is defined by
\begin{eqnarray}
\label{eq1}
\Gamma(I\!\!\!\!C) = \bigoplus_{n=0}^{\infty}I\!\!\!\!C |n\rangle
\end{eqnarray}
for any $n \in I\!\!\!\!N$ where $I\!\!\!\!C |n\rangle$ is called the $n$-particle subspace. The different $n$-particle subspaces are
orthogonal, that is, the sum in (\ref{eq1}) is orthogonal. The square of the semi-norm of the vector $|n\rangle$ is given by
\begin{equation}
\label{eq2}
\langle n|n\rangle = \lambda_{n}
\end{equation}
where $\lambda_{n}$ being a real number and $\lambda_{n} \ge 0$ for each $n\in I\!\!\!\!N$ and if for some $n$ we have
$\lambda_{n} = 0$, then $\lambda_{m} = 0$ for all $m \ge n$. After taking quotient, the semi-norm in (\ref{eq2}) becomes a norm which makes $\Gamma(I\!\!\!\!C)$ a
pre-Hilbert space. In the following we will consider its completion, which, with an abuse of notation, will be denoted by $\Gamma(I\!\!\!\!C)$.

An arbitrary vector $f$ in $\Gamma(I\!\!\!\!C)$ is given by
\begin{eqnarray}\nonumber
f \equiv c_{0}|0\rangle + c_{1}|1\rangle + c_{2}|2\rangle + \ldots + c_{n}|n\rangle +
\ldots
\end{eqnarray}
for any $n \in I\!\!\!\!N$ with $\|f\| = ( \sum_{n=0}^{\infty}
|c_{n}|^{2} \lambda_{n} )^{1/2} < \infty.$

We now consider the following actions on $\Gamma(I\!\!\!\!C)$ :
\begin{eqnarray}
  \begin{array}{rcl}\nonumber
    A^{\dag} |n\rangle & = & |n+1\rangle\\\nonumber
    A|n+1\rangle & = & \frac{\lambda_{n+1}}{\lambda_{n}} |n\rangle
  \end{array}
\end{eqnarray}
$A^{\dag}$ is called the {\em creation operator} and its adjoint $A$ is called the {\em annihilation operator}.

The commutation relation takes the form
\begin{eqnarray}\nonumber
[A, A^{\dag}] = \frac{\lambda_{N+1}}{\lambda_{N}} -
\frac{\lambda_{N}} {\lambda_{N-1}}
\end{eqnarray}
where $N$ is the number operator defined by $N|n\rangle = n|n\rangle$.

In a paper \cite{das2}, we have proved that the set $\left\{\left|\frac{n}{\sqrt{\lambda_{n}}}\right\rangle, n = 0, 1, 2, 3, \ldots \right\}$ forms a complete
orthonormal set and the solution of the following eigenvalue equation
\begin{eqnarray}\nonumber
Af_{\alpha} = \alpha f_{\alpha}
\end{eqnarray}
is given by
\begin{eqnarray}\nonumber
f_{\alpha} = \psi(|\alpha|^{2})^{-1/2} \sum_{n=0}^{\infty} \frac{\alpha^{n}}{\lambda_{n}}|n\rangle
\end{eqnarray}
where $\psi(|\alpha|^{2}) = \sum_{n=0}^{\infty} \frac{|\alpha|^{2n}} {\lambda_{n}}$. We call $f_{\alpha}$ a {\bf coherent vector} in $\Gamma(I\!\!\!\!C).$

Now, we observe that
$$AA^{\dag} = \frac{\lambda_{N+1}}{\lambda_{N}},\:\:\:\:
A^{\dag}A = \frac{\lambda_{N}}{\lambda_{N-1}} $$

We further observe that $\left(\frac{\lambda_{N+1}}{\lambda_{N}} - \frac{\lambda_{N}} {\lambda_{N-1}}\right)$ commutes with both $A^{\dag}A$ and $AA^{\dag}.$

\section{Time Evolution of State Vector}

The scheme of the $V$-type three-level atomic system consists of two allowed transitions $$|a\rangle\leftrightarrow|c\rangle\:\:\mbox{and}\:\:
|b\rangle \leftrightarrow |c\rangle$$ where $|a\rangle$, $|b\rangle$ and $|c\rangle$ are excited state, intermediate state and ground state respectively. Each interaction has a different mode of the field. In the rotating-wave approximation, its Hamiltonian is described by
\begin{eqnarray}
\label{eq3}
H = H_{0} + H_{1},
\end{eqnarray}
where taking $\hbar=1$,
\begin{eqnarray*}
H_{0} = \omega_{a}|a\rangle\langle a| + \omega_{b}|b\rangle\langle b|
+ \omega_{c}|c\rangle\langle c| +\gamma A^{\dag}A ,
\end{eqnarray*}
and
\begin{eqnarray*}
H_{1} = g_{1}A|a\rangle\langle c| + g_{1}A^{\dag}|c\rangle\langle a|+ g_{2}A|b\rangle\langle c| + g_{2}A^{\dag}|c\rangle\langle b|.
\end{eqnarray*}
Here $A^{\dag}$ and $A$ are, respectively, the creation and annihilation operators for the field of frequency $\gamma$. $|i\rangle(i=a, b, c)$ is the eigenstate  of the atom with eigenfrequency $\omega_{i}$ and $g_1,\,g_2$ are the corresponding coupling constant. We assume the coupling constants to be real
throughout the paper.

\begin{figure*}
\begin{center}
\includegraphics[width=3.5in,height=1.5in]{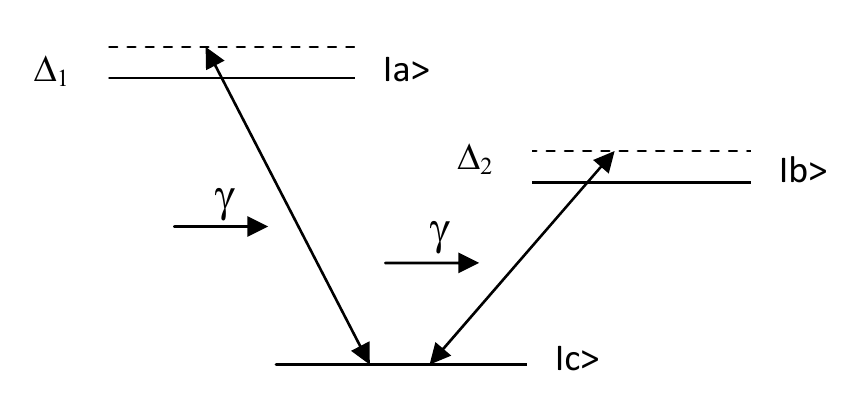}
\caption{Energy diagram of a $V$-shaped three-level
atom interacting with one quantized cavity mode.}
\label{fig:1}       
\end{center}
\end{figure*}

In the interaction picture, the state vector of this atom-field coupling system at time $t$ can be described by
\begin{eqnarray}
\begin{array}{rcl}
& & |\psi(t)\rangle\\
& = & \sum_{n}\left(C_{a, n-1}\left|a,
\frac{n-1}{\sqrt{\lambda_{n-1}}}\right\rangle +\, C_{b, n-1}\left|b,
\frac{n-1}{\sqrt{\lambda_{n-1}}}\right\rangle +\, C_{c,
n}\left|c, \frac{n}{\sqrt{\lambda_{n}}}\right\rangle\right)
\label{eq4.1}
\end{array}
\end{eqnarray}

In the interaction picture, the Hamiltonian (\ref{eq3}) is given by
\begin{eqnarray}
\begin{array}{rcl}
V & = & g_{1}e^{i\triangle_{1} t}A|a\rangle\langle c|+
g_{1}A^{\dag}e^{-i\triangle_{1} t}|c\rangle\langle a| + g_{2}e^{i\triangle_{2} t}A|b\rangle\langle c|\\\\
& & +g_{2}A^{\dag}e^{-i\triangle_{2} t}|c\rangle\langle b|
\label{eq5.1}
\end{array}
\end{eqnarray}
where
\begin{eqnarray*}
\begin{array}{lcl}
\Delta_{1} & = & \omega_{a} - \omega_{c} - \gamma\left(
\frac{\lambda_{n+1}}{\lambda_{n}} - \frac{\lambda_{n}}{\lambda_{n-1}}\right),\\\\
\Delta_{2} & = & \omega_{b} - \omega_{c} - \gamma\left(\frac{\lambda_{n+1}}{\lambda_{n}} -
\frac{\lambda_{n}}{\lambda_{n-1}}\right).
\end{array}
\end{eqnarray*}

On solving the Schr\"{o}dinger equation $i\hbar \frac{\partial}{\partial t}|\psi(t)\rangle=V|\psi(t)\rangle$ with help of (\ref{eq4.1}) and (\ref{eq5.1}) and assuming $\Delta_1=\Delta_2=\Delta'$, we get the equations of motion for
probability amplitudes as
\begin{eqnarray}
\left.
\begin{array}{lcl}
\dot{C}_{a, n-1}  & = & -i g_{1}\sqrt{\frac{\lambda_{n}}{\lambda_{n-1}}}
e^{i\triangle^{'}t}C_{c, n},\\\\
\dot{C}_{b,n-1}  & = &
-i g_{2}\sqrt{\frac{\lambda_{n}}{\lambda_{n-1}}}
e^{i\triangle^{'}t}C_{c, n},\\\\
\dot{C}_{c, n} & = & -i g_{1}\sqrt{\frac{\lambda_{n}}{\lambda_{n-1}}}e^{-i\triangle^{'}t}C_{a, n-1} -i g_{2}\sqrt{\frac{\lambda_{n}}{\lambda_{n-1}}}
e^{-i\triangle^{'}t}C_{b,n-1}
\label{eq6}
\end{array}
\right\}
\end{eqnarray}
where we assume
\begin{eqnarray*}
\begin{array}{lcl}
\Delta^{'} & = & \omega_{a} - \omega_{c} - \gamma\left(\frac{\lambda_{n+1}}{\lambda_{n}} -
\frac{\lambda_{n}}{\lambda_{n-1}}\right)\\\\
& = & \omega_{b} - \omega_{c} - \gamma\left(\frac{\lambda_{n+1}}{\lambda_{n}} -
\frac{\lambda_{n}}{\lambda_{n-1}}\right)
\end{array}
\end{eqnarray*}

If the atom is initially in the state $|\psi_{a}(0)\rangle = \cos \frac{\alpha}{2}|a\rangle + \sin \frac{\alpha}{2}e^{-i\psi}|b\rangle$ which means that the atom is in the coherent superposition state of its eigenkets $|a\rangle$ and $|b\rangle$, and the field is in the superposition of the photon number states at time $t=0$,
$|\psi_{f}(0)\rangle = \sum_{n}F_{n}|\frac{n}{\sqrt{\lambda_{n}}}\rangle$ with $\sum_{n}|F_{n}|^{2} = 1$, then the state vector of the total system at $t=0$ can be described as
\begin{eqnarray*}
|\psi(0)\rangle  =  \sum_{n}\left(\cos \frac{\alpha}{2}F_{n-1}\left|a, \frac{n-1}{\sqrt{\lambda_{n-1}}}\right\rangle+ \sin \frac{\alpha}{2}
e^{-i\psi}F_{n-1}\left|b,\frac{n-1}{\sqrt{\lambda_{n-1}}}\right\rangle\right)
\end{eqnarray*}

With this initial condition, solving (\ref{eq6}) we get
\begin{eqnarray*}
C_{c, n}(t)  =  B_{1}\{e^{-i(\triangle^{'}/2 + \beta)t} - e^{-i(\triangle^{'}/2 - \beta)t}\}
\end{eqnarray*}
where
$$B_{1} = \frac{g_{1}\sqrt{\frac{\lambda_{n}}{\lambda_{n-1}}}\cos
\frac{\alpha}{2}F_{n}\\
+ g_{2}\sqrt{\frac{\lambda_{n}}
{\lambda_{n-1}}}\sin \frac{\alpha}{2}e^{-i\psi}F_{n}}{2\beta}$$
and
$$\beta^{2} = {\Delta^{'}}^{2}/4 + (g_{1}^{2}
+g_{2}^{2})\frac{\lambda_{n}}{\lambda_{n-1}}$$
where $\beta$ is associated with the frequency of the atomic Rabi oscillation. Similarly we get
\begin{eqnarray*}
C_{a, n-1}(t) =
 -g_{1}\sqrt{\frac{\lambda_{n}}{\lambda_{n-1}}}
B_{1}\left[\frac{e^{i(\Delta^{'}/2+\beta) t} - 1
}{(\Delta^{'}/2+\beta)} - \frac{e^{i(\Delta^{'}/2-\beta) t} - 1
}{(\Delta^{'}/2-\beta)}\right]+ \cos \frac{\alpha}{2}F_{n-1}
\end{eqnarray*}
and
\begin{eqnarray*}
C_{b,n-1}(t)= -g_{2}\sqrt{\frac{\lambda_{n}}
{\lambda_{n-1}}}B_{1}\left[\frac{e^{i(\Delta^{'}/2+\beta) t} - 1
}{(\Delta^{'}/2+\beta)} - \frac{e^{i(\Delta^{'}/2-\beta) t} - 1
}{(\Delta^{'}/2-\beta)}\right] + \sin \frac{\alpha}{2}e^{-i\psi}F_{n-1}
\end{eqnarray*}
Substituting the values of $C_{c, n}(t)$, $C_{a, n-1}(t)$ and
$C_{b,n-1}(t)$ in (\ref{eq4.1}) we can obtain the state vector of the system at time $t$ in the interaction picture.

At this stage we assume that $$\alpha = 90^{0}\,\,\mbox{and}\,\,\psi =0,\,\,\mbox{Also}\,\,F_{n} \approx F_{n-1}$$
This reduces the coefficients of (\ref{eq4.1}) into

\begin{eqnarray}
\left.
\begin{array}{rcl}
C_{c, n}(t)
& = & B_1 \{e^{-i({\triangle}'/2 + \beta)t} - e^{-i({\triangle}'
/2 - \beta)t}\}\\\\
C_{a, n-1}(t) & = & -g_{1}\sqrt{\frac{\lambda_{n}}{\lambda_{n-1}}}
B_{1}\left[\frac{e^{i(\Delta^{'}/2+\beta) t} - 1
}{(\Delta^{'}/2+\beta)} - \frac{e^{i(\Delta^{'}/2-\beta)
t} - 1 }{(\Delta^{'}/2-\beta)}\right]+ \frac{1}{\sqrt{2}}F_{n}\\\\
C_{b,n-1}(t) & = & -g_{2}\sqrt{\frac{\lambda_{n}}
{\lambda_{n-1}}}B_{1}\left[\frac{e^{i(\Delta^{'}/2+\beta) t} -
1 }{(\Delta^{'}/2+\beta)} - \frac{e^{i(\Delta^{'}/2-\beta)
t} - 1 }{(\Delta^{'}/2-\beta)}\right]+ \frac{1}{\sqrt{2}}F_{n}
\end{array}
\right\}
\end{eqnarray}
with
$$
B_{1} = \sqrt{\frac{\lambda_n}{\lambda_{n-1}}}\frac{F_{n}(g_{1}+g_{2})}{2\sqrt{2}\beta}\,\,\,\,\mbox{and}\,\,\,\,
\beta^{2} = {\Delta^{'}}^{2}/4 + (g_{1}^{2}
+g_{2}^{2})\frac{\lambda_{n}}{\lambda_{n-1}}
$$

We assume that the atom enters the cavity with the initial state
\begin{eqnarray*}
|\psi(0)\rangle =  \sum_{n} \frac{1}{\sqrt{2}}F_{n-1}\left(\left|a,
\frac{n-1}{\sqrt{\lambda_{n-1}}}\right\rangle + \left|b,
\frac{n-1}{\sqrt{\lambda_{n-1}}}\right\rangle\right)
\end{eqnarray*}
and after the evolution for time $t_{1}$, the state vector of the considered atom-field system becomes
\begin{eqnarray}
\begin{array}{rcl}
\label{eq8}
|\psi(t_{1})\rangle & = &  \sum_{n}\left[C_{a, n-1}(t_{1})\left|a,
\frac{n-1}{\sqrt{\lambda_{n-1}}}\right\rangle + C_{b, n-1}(t_{1})\left|b,
\frac{n-1}{\sqrt{\lambda_{n-1}}}\right\rangle\right.\\
& &  \left.+ C_{c, n}(t_{1})\left|c,\frac{n}{\sqrt{\lambda_{n}}}\right\rangle\right]
\end{array}
\end{eqnarray}


Further assuming $g_{1}= g_{2} = g$ with zero detuning and $\Delta^{'} = 0$, the system evolves to $|\psi(t_{1})\rangle$, given by (\ref{eq8}),
where
\begin{eqnarray}
\left.
\begin{array}{rcl}
C_{c, n}(t_{1}) & = & -iF_{n}\sin \beta t_{1}\\\\
C_{a, n-1}(t_{1}) & = &-\frac{1}{\sqrt{2}}F_{n}(\cos \beta t_{1} -2)\\\\
C_{b,n-1}(t_{1}) & = & -\frac{1}{\sqrt{2}}F_{n}(\cos \beta t_{1} -2)
\label{eq9}
\end{array}
\right\}
\end{eqnarray}

The state vector $|\psi(t_{1})\rangle$ describes the time evolution of the whole atom-field system but we now concentrate on
some statistical properties of the single-mode cavity field. The field inside the cavity after departing the  atom is obtained by
tracing out the atomic part of $|\psi(t_{1})\rangle$ as
\begin{equation}
\label{eq10}
{|\psi(t_{1})\rangle}_{f} = Tr_{a}[|\psi(t_{1})\rangle],
\end{equation}
where we have used the subscript $a\,(f)$ to denote the atom (field).

This ${|\psi(t_{1})\rangle}_{f}$ will be of consideration throughout the next section to determine the statistical properties of the field left into the cavity.


\section{Statistical properties of the radiation field}

In this section we investigate about two nonclassical effects, namely, sub-Poissonian photon statistics
and quadrature squeezing of the radiation field.

\subsection{Sub-Poissonian photon statistics}

The Mandel parameter $Q^M$ illustrates the nonclassicality through photon number distribution of a quantum state \cite{peng2,mandel}. It is defined as
\begin{equation}
\label{eq7}
Q^{M} \equiv \frac{\langle n^{(2)}\rangle}{\langle n\rangle} -
\langle n\rangle
\end{equation}
where
$$\langle n^{(2)}\rangle = {}_{f}\langle\psi(t_1)|A^{\dag}A^{\dag}AA|\psi(t_1)\rangle_f\,\,\mbox{and}\,\,\langle n\rangle = {}_{f}\langle\psi(t_1)|A^{\dag}A|\psi(t_1)\rangle_f.$$ The negative values of $Q^M$ parameter essentially indicate the negativity for $P$ function and so it gives a witness for nonclassicality. For the Poissonian statistics it becomes 0, while for the sub-Poissonian (super-Poissonian) photon statistics it has negative (positive) values.

Using
\begin{eqnarray*}
A^{\dag}A\left|\frac{n}{\sqrt{\lambda_{n}}}\right\rangle  =  \frac{\lambda_{n}}{\lambda_{n-1}}
\left|\frac{n}{\sqrt{\lambda_{n}}}\right\rangle,
\end{eqnarray*}
and (\ref{eq10}), we have calculated the analytical expressions for the first and second order moments as
\begin{eqnarray*}
\langle A^{\dag}A\rangle = \sum_{n}2C_{a,n-1}(t_{1})\bar{C}_{a,n-1}(t_{1})
\frac{\lambda_{n-1}}{\lambda_{n-2}}+\sum_{n}C_{c,n}(t_{1})\bar{C}_{c, n}(t_{1})\frac{\lambda_{n}}{\lambda_{n-1}}
\end{eqnarray*}
and
\begin{eqnarray*}
\langle A^{\dag}A^{\dag}AA\rangle
= \sum_{n}2C_{a,n-1}(t_{1})\bar{C}_{a,n-1}(t_{1})
\frac{\lambda_{n-1}}{\lambda_{n-3}}+\sum_{n}C_{c,
n}(t_{1})\bar{C}_{c, n}(t_{1})\frac{\lambda_{n}}{\lambda_{n-2}}
\end{eqnarray*}

Thus
\begin{eqnarray*}
Q^{M} & = & \frac{\langle A^{\dag}A^{\dag}AA\rangle}{\langle A^{\dag}A\rangle}
- \langle A^{\dag}A\rangle\\
& = & \frac{\sum_{n}2C_{a,n-1}(t_{1})\bar{C}_{a,n-1}(t_{1})
\frac{\lambda_{n-1}}{\lambda_{n-3}}  +\sum_{n}C_{c,
n}(t_{1})\bar{C}_{c, n}(t_{1})\frac{\lambda_{n}}{\lambda_{n-2}}}{
\sum_{n}2C_{a,n-1}(t_{1})\bar{C}_{a,n-1}(t_{1})
\frac{\lambda_{n-1}}{\lambda_{n-2}}  +\sum_{n}C_{c,
n}(t_{1})\bar{C}_{c, n}(t_{1})\frac{\lambda_{n}}{\lambda_{n-1}}}\\
&   & -\sum_{n}2C_{a,n-1}(t_{1})\bar{C}_{a,n-1}(t_{1})
\frac{\lambda_{n-1}}{\lambda_{n-2}}-\sum_{n}C_{c,
n}(t_{1})\bar{C}_{c, n}(t_{1})\frac{\lambda_{n}}{\lambda_{n-1}}
\end{eqnarray*}

Substituting $C_{c,n}(t_{1})$ and $C_{a,n-1}(t_{1})$ from (\ref{eq9}), we get

\begin{eqnarray*}
Q^{M}= \frac{A+B}{C+D}-\left[\sum_{n}|F_{n}|^{2}(\cos \beta t_{1}-2)^{2}
\frac{\lambda_{n-1}}{\lambda_{n-2}} +\sum_{n}|F_{n}|^{2}\sin^{2}\beta t_{1}
\frac{\lambda_{n}}{\lambda_{n-1}}\right]
\end{eqnarray*}
with
\begin{eqnarray*}
\begin{array}{rcl}
A & = & \sum_{n}|F_{n}|^{2}(\cos \beta t_{1}-2)^{2}
\frac{\lambda_{n-1}}{\lambda_{n-3}}\\\\
B & = & \sum_{n}|F_{n}|^{2}\sin^{2}\beta t_{1}
\frac{\lambda_{n}}{\lambda_{n-2}}\\\\
C & = & \sum_{n}|F_{n}|^{2}(\cos \beta t_{1} -2)^{2}
\frac{\lambda_{n-1}}{\lambda_{n-2}}\\\\
D & = & \sum_{n}|F_{n}|^{2}\sin^{2}\beta t_{1}
\frac{\lambda_{n}}{\lambda_{n-1}}
\end{array}
\end{eqnarray*}
If the radiation field is initially in a coherent state \cite{peng1}, then
$F_{n}(0)=\exp(-\bar{n}/2)\frac{\bar{n}^{n/2}e^{i\zeta n}}{\sqrt{n!}}$. Substituting $F_{n}(0)$, assuming
$\beta t_{1}\equiv \theta_{1}$ and finally taking $t_{1}=t$ so that $\theta_{1}=\theta$ with
$\theta=\sqrt{2\frac{\lambda_{n}}{\lambda_{n-1}}}gt$, we get

\begin{eqnarray}
Q^{M} = \frac{A^{'}+B^{'}}{C^{'}+D^{'}}-e^{-\bar{n}}\left[\sum_{n}\frac{\lambda_{n-1}}{\lambda_{n-2}}
\frac{\bar{n}^{n}}{n!}(\cos \theta -2)^{2}+\sum_{n}\frac{\lambda_{n}}{\lambda_{n-1}}
\frac{\bar{n}^{n}}{n!}(\sin\theta)^2\right]
\label{eq12}
\end{eqnarray}
where
\begin{eqnarray*}
\begin{array}{rcl}
A^{'} & = & e^{-\bar{n}}\sum_{n}\frac{\lambda_{n-1}}
{\lambda_{n-3}}\frac{\bar{n}^{n}}{n!}(\cos \theta -2)^{2}\\\\
B^{'} & = & e^{-\bar{n}}\sum_{n}\frac{\lambda_{n}}{\lambda_{n-2}}
\frac{\bar{n}^{n}}{n!}(\sin\theta)^2\\\\
C^{'} & = & e^{-\bar{n}}\sum_{n}\frac{\lambda_{n-1}}{\lambda_{n-2}}
\frac{\bar{n}^{n}}{n!}(\cos \theta -2)^{2}\\\\
D^{'} & = & e^{-\bar{n}}\sum_{n}\frac{\lambda_{n}}{\lambda_{n-1}}
\frac{\bar{n}^{n}}{n!}(\sin\theta)^2
\end{array}
\end{eqnarray*}

\begin{figure*}[ht]
\centering
\includegraphics[width=5cm]{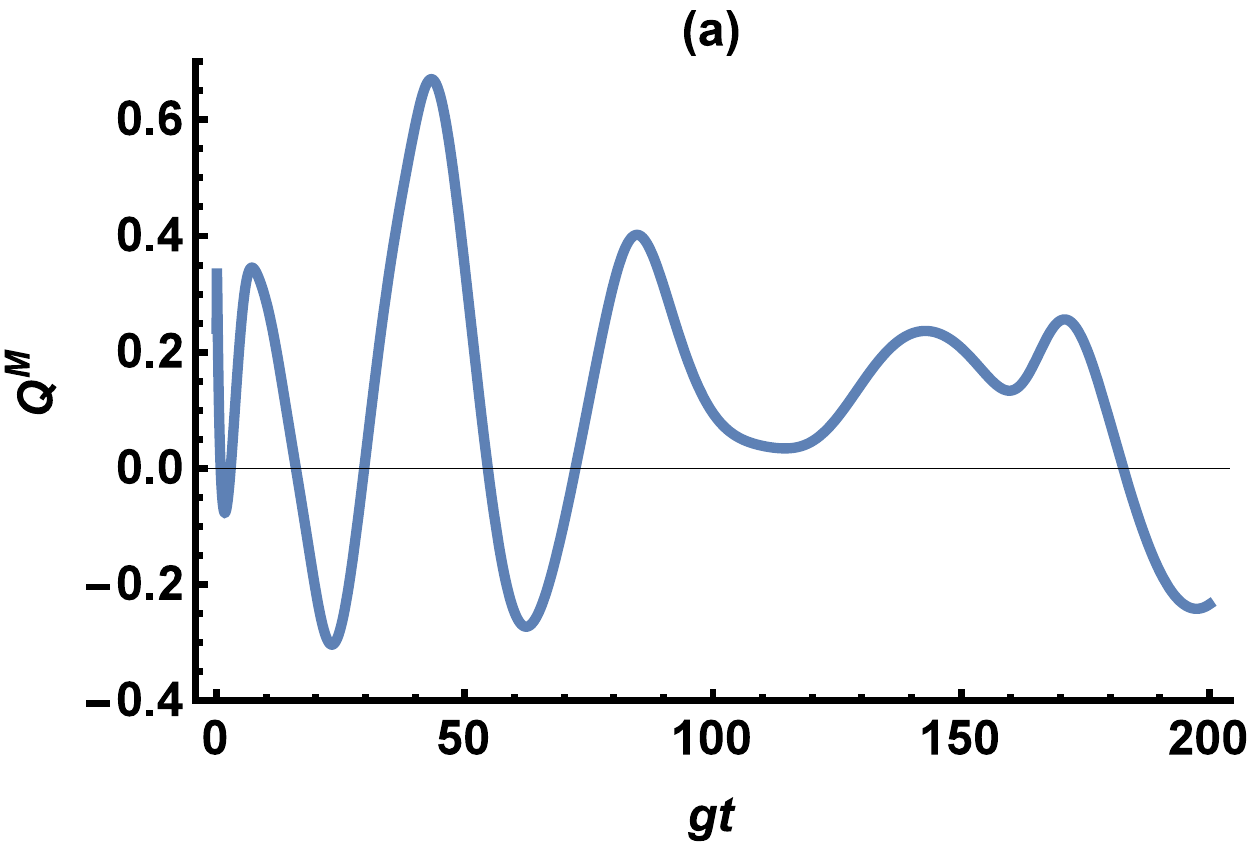}\hspace{1cm}
\includegraphics[width=5cm]{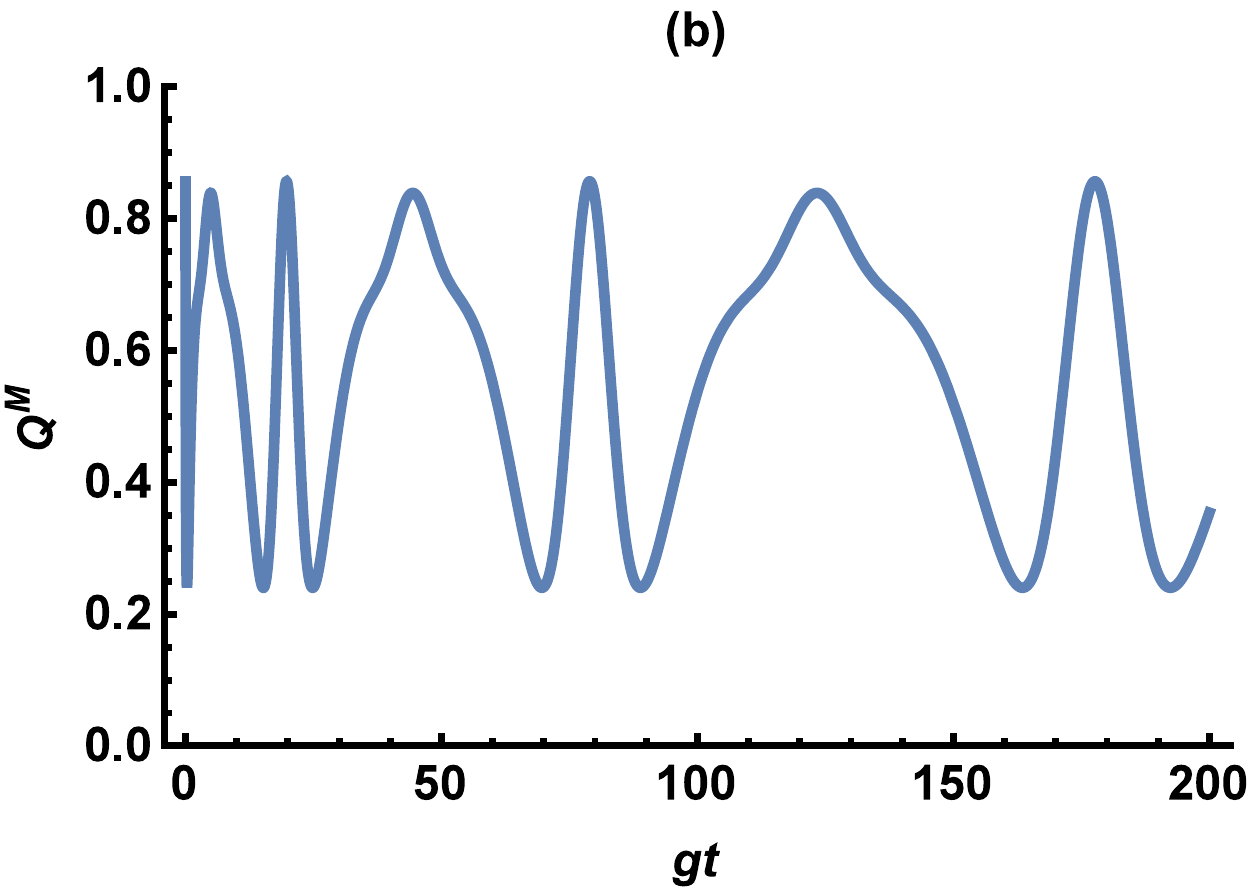}\hspace{1cm}
\includegraphics[width=5cm]{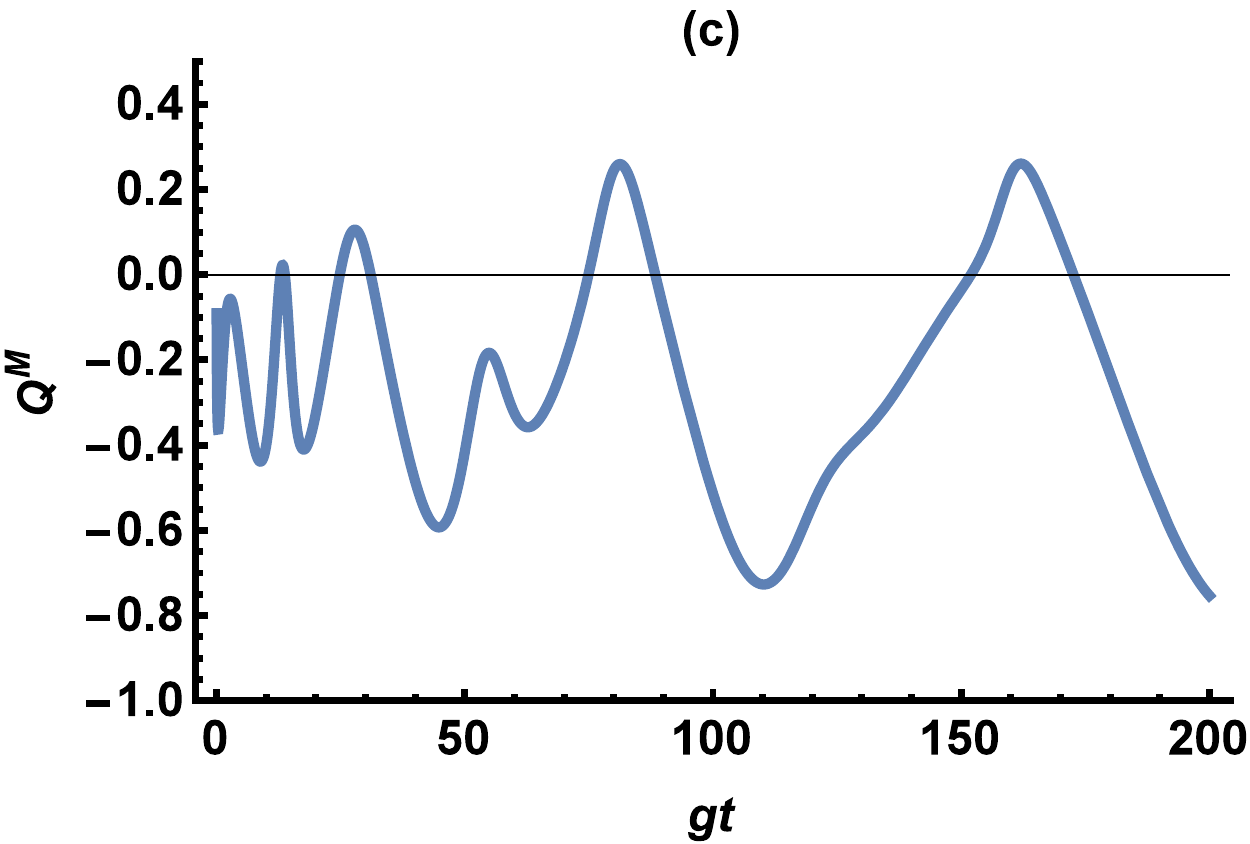}\hspace{1cm}

\caption{(Color online) Mandel's $Q^M$ as a function of $gt$ for $\bar{n}=0.5$ and for different values of $\lambda_n$ such that $(a)~\lambda_n=n!$; $(b)~\lambda_n=(n!)^2$ and $(c)~\lambda_n=[n]$.}
\label{fig2}
\end{figure*}
In Fig.~\ref{fig2}, the dependence of $Q^M$ using (\ref{eq12}) is shown over the scaled time $gt$, for an initial single-mode coherent field with $\bar{n} = 0.5$. For simplicity, we have considered $g_1 = g_2 = g$ in all of our numerical calculations. The negative values of $Q^M$ parameter essentially indicate the negativity of the $P$ function and hence it represents a witness for nonclassicality.  When $\lambda_n = n!$, $Q^M$ parameter fluctuates between certain positive and negative values, which illustrates the existence of sub-Poissonian nature of the state. For $gt\approx 25$, the cavity field attains the most negative $Q^M$ value ($\approx -0.3$), demonstrating the nonclassical behaviour of the radiation field [cf. Fig.\,\ref{fig2}(a)]. For $\lambda_n = (n!)^2$, $Q^M$ function oscillates but always remains positive. In this case, Mandel's parameter fails to identify the nonclassicality property of the cavity field. When $\lambda_n$ takes the value $[n] = \frac{1-q^n}{1-q}$, $q=0.5$, $Q^M$ almost settles to the sub-Poissonian statistics. That means, the nonclassical behaviour of the radiation field depends on the choice of $\lambda_n$.

So far we have considered a basic model which consists of a $V$-type three level atom passing through a high-$Q$ cavity, interacting with a single-mode coherent field contained in that cavity. Although this simple atom and single photonic mode system is powerful and intuitive, it is quite challenging to realize in practice \cite{cohen}. Because, in reality, there are typically numerous modes that all interact with an atom. All of these modes make up a continuous spectrum of vacuum fluctuations that all attempt to develop an excited atomic Rabi flopping. The resulting sum of all of the continuums of probability amplitudes that interfere gives rise to the more commonly observed exponential decay of an excited atom by dissipating the energy irreversibly into its environment. Thus, to demonstrate a system where the coupling strength between the atom and a particularly chosen single photonic mode (e.g., single mode defined by an optical fiber) is much stronger relative to all other dissipative channels such as the vacuum/environment, is of great challenge. Also, when a radiation field propagates through the environment, it inevitably interacts with its surroundings, which
causes the decoherence \cite{arpita3}. Since it is impossible to perfectly isolate a quantum system from its environment, decoherence effects are more or less unavoidable. It is well known that decoherence will deteriorate the degree of nonclassicality of the optical fields \cite{sb}. Here, we introduce only the cavity decay and study the effect of loss of cavity over the nonclasscality witnesses under consideration. Assuming that there is no photon leakage from the cavity, the evolution of the system is governed by the non-Hermitian Hamiltonian \cite{deng,arpita2}
\begin{eqnarray}
H_{\mathrm{loss}} = H-\frac{ik}{2} A^\dagger A,
\end{eqnarray}
where $k$ is the cavity decay rate and $H$ is given by (\ref{eq3}). Solving Schr\"{o}dinger equation with respect to $H_{\mathrm{loss}}$ and assuming $\Delta'=0$, $g_1=g_2=g$, the state vector of this atom-field coupled system at time $t_2$ can be described by $|\psi'(t_2)\rangle$ with
\begin{eqnarray}
\left.
\begin{array}{rcl}
\label{eq14}
C'_{c, n}(t_2)
& = & -2B_1 e^{-\frac{kt_2}{4}}\sin{\beta' t_2}\\\\
C'_{a, n-1}(t_2) & = & -g \sqrt{\frac{\lambda_{n}}{\lambda_{n-1}}}
B_{1}\left[\frac{1}{\beta'^2+\frac{k^2}{4}}(\cos{\beta' t_2}-2)+\frac{k^2}{\beta'^2+\frac{k^2}{4}}\sin{\beta' t_2}\right]\\\\
& & + \frac{1}{\sqrt{2}}F_{n}\\\\
C'_{b, n-1}(t_2) & = & -g \sqrt{\frac{\lambda_{n}}{\lambda_{n-1}}}
B_{1}\left[\frac{1}{\beta'^2+\frac{k^2}{4}}(\cos{\beta' t_2}-2)+\frac{k^2}{\beta'^2+\frac{k^2}{4}}\sin{\beta' t_2}\right]\\\\
& & + \frac{1}{\sqrt{2}}F_{n}
\end{array}
\right\}
\end{eqnarray}
with $$B_1 = \sqrt{\frac{\lambda_n}{\lambda_{n-1}}}\frac{F_n\,g}{\sqrt{2}\beta'}\,\,\,\,\mathrm{and}\,\,\,\,\beta'^2 = 2g^2\left(\frac{\lambda_n}{\lambda_{n-1}}\right)-\frac{k^2}{16}$$
By using $|\psi'(t_2)\rangle$ given in (\ref{eq14}), the Mandel's parameter after taking into account the loss of cavity, $Q^M_{\mathrm{loss}}$, can be easily obtained from (\ref{eq7}).
\begin{figure*}[ht]
\centering

\includegraphics[width=5cm]{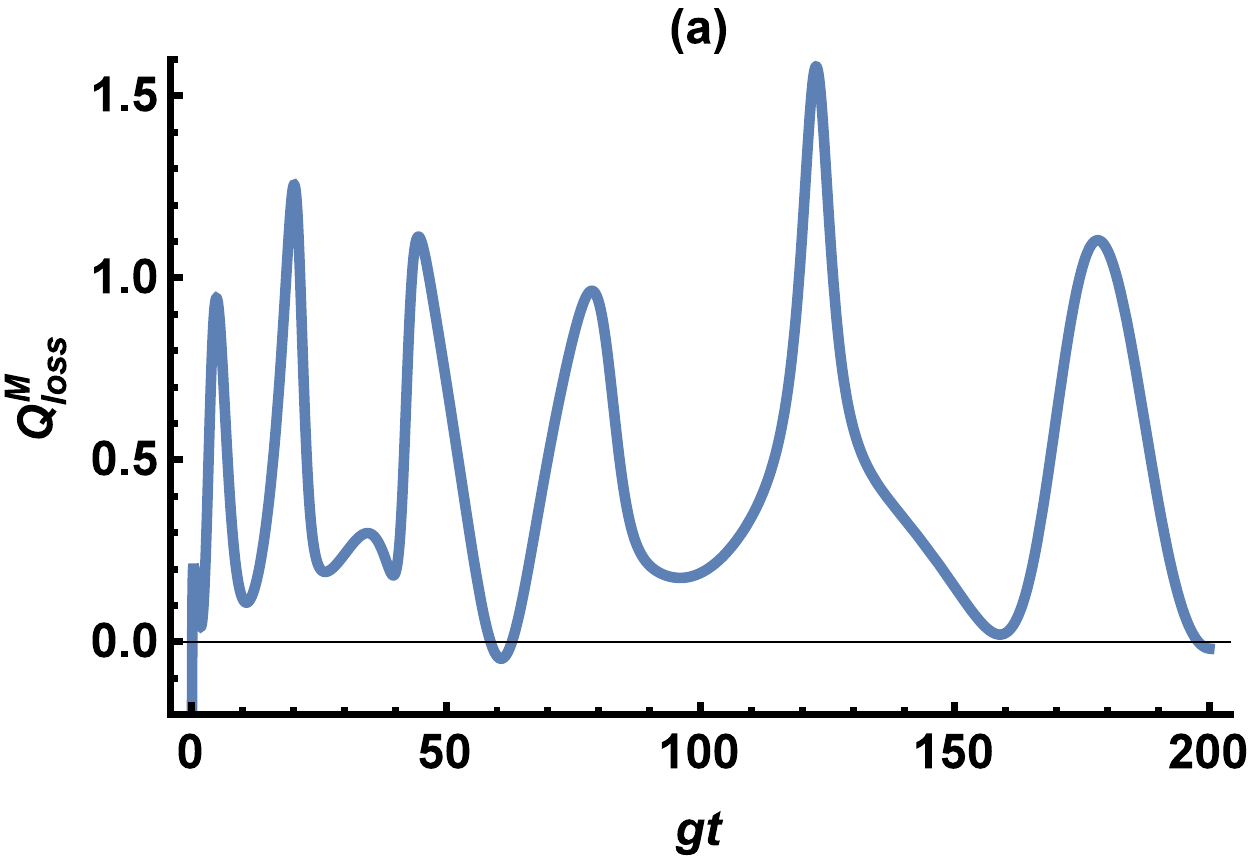}\hspace{1cm}
\includegraphics[width=5cm]{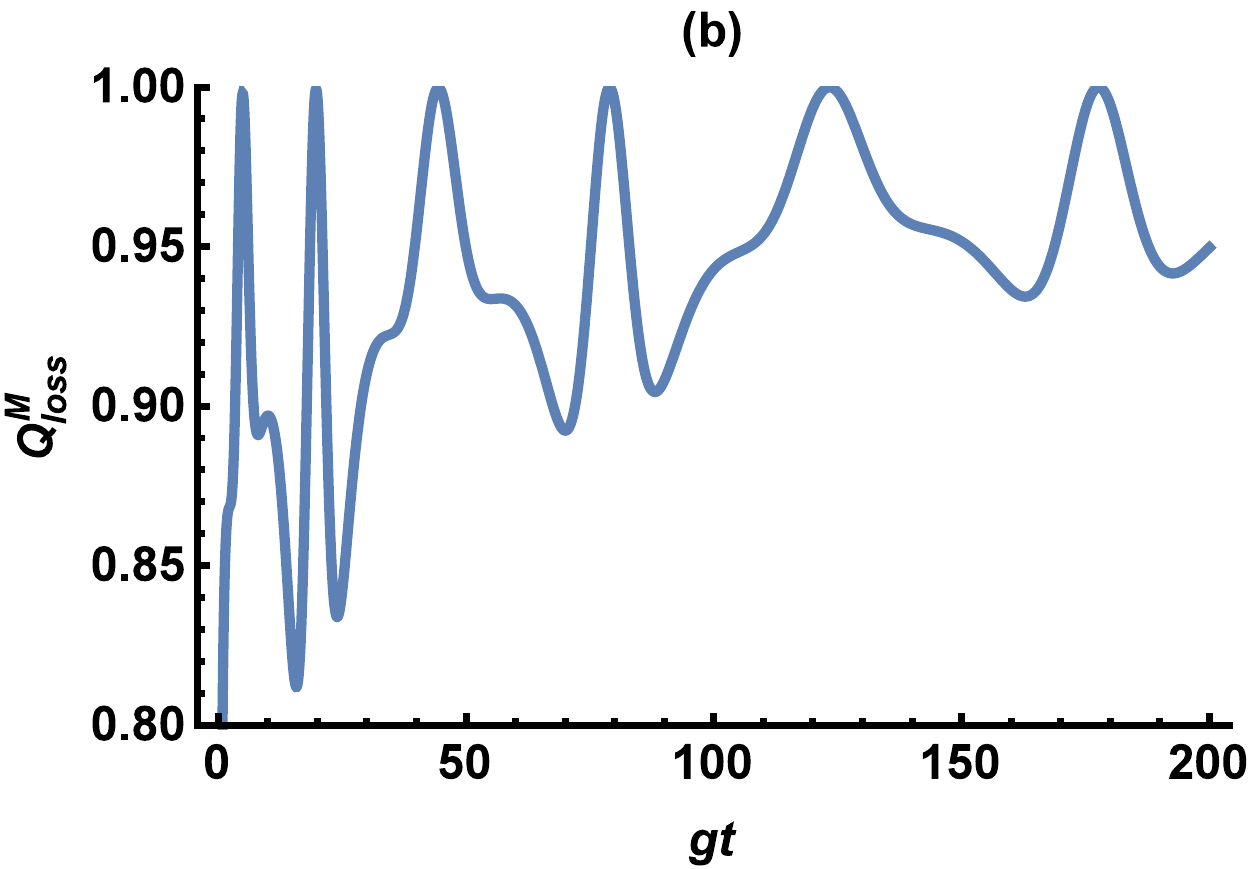}\hspace{1cm}
\includegraphics[width=5cm]{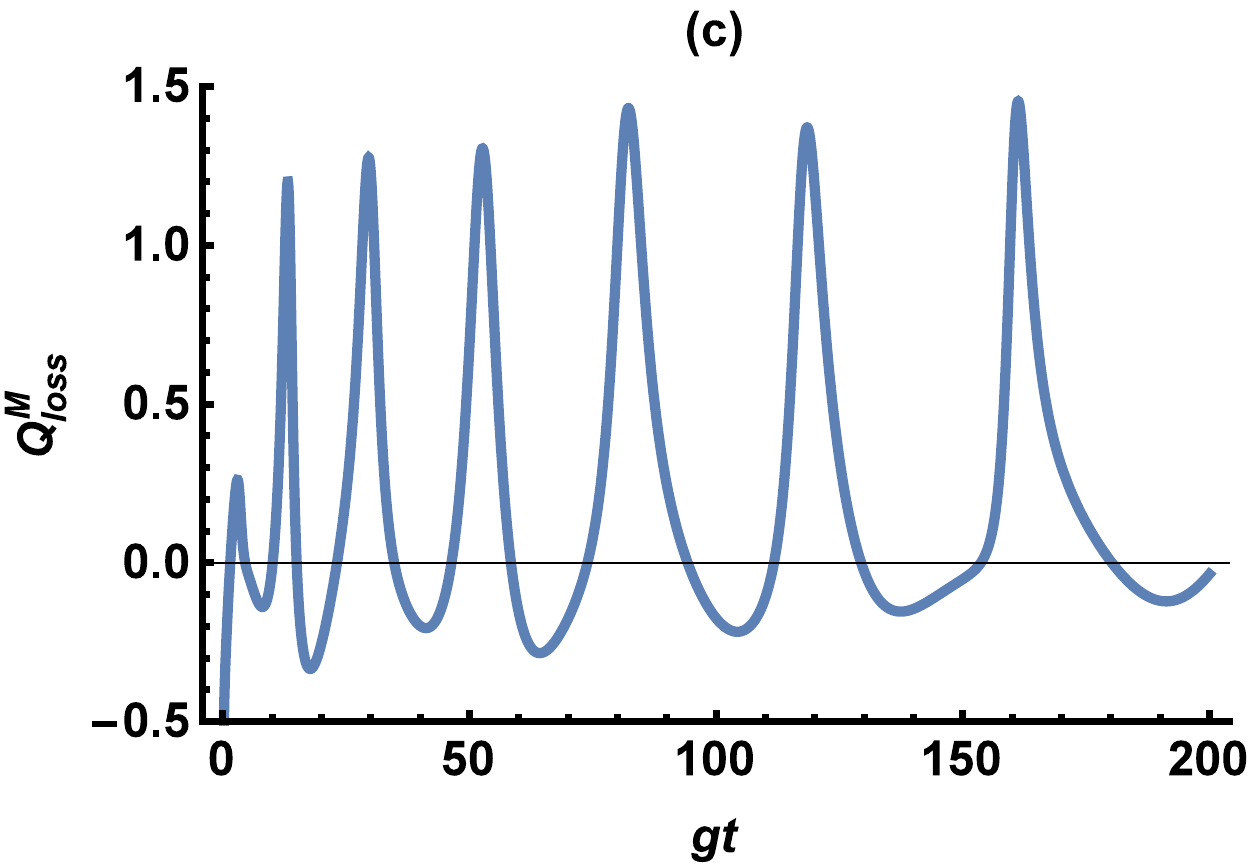}\hspace{1cm}

\caption{(Color online) Mandel's $Q^M$ after loss as a function of $gt$ for $\bar{n}=0.5$, $k=0.1$ and for different values of $\lambda_n$ such that $(a)~\lambda_n=n!$; $(b)~\lambda_n=(n!)^2$ and $(c)~\lambda_n=[n]$.}
\label{fig3}
\end{figure*}
In Fig.~\ref{fig3}, we have shown the variation of $Q^M_{\mathrm{loss}}$ for cavity decay rate $k=0.1$. Other parametric values remain same as in Fig.~\ref{fig2}. We have noticed that for all varieties of $\lambda_n$ such as $n!,~(n!)^2,~[n]$, the Mandel's parameter approaches non-negative values more quickly than without decay case. That means the presence of cavity decay causes a notable change in decreasing the nonclassicality of the radiation field. It is of further investigation that how nonclassicality of the cavity field  is affected by higher values of $k$.

\subsection{Squeezing properties of the radiation field}

A general class of minimum-uncertainty states are known as squeezed states \cite{walls}. A squeezed may have less noise in one quadrature than a coherent state and thus to satisfy the requirements for being a minimum-uncertainty state, the noise in the other quadrature is greater than that of a coherent state. That means, if the fluctuation of the radiation field $\triangle X$ in a quadrature $X$ goes below the square root of the uncertainty product, the fluctuation $\triangle Y$ in the other quadrature $Y$ should be greater than that of and vice-versa \cite{girish}. To analyze the squeezing properties of the radiation field \cite{lee2,wang1}, we introduce two hermitian quadrature operators
\begin{eqnarray*}
X=A+A^{\dag},\,\,\,\,\,Y= -i(A-A^{\dag})
\end{eqnarray*}
These two quadrature operators satisfy the commutation relation
\begin{eqnarray*}
[X, Y]=2i\left(\frac{\lambda_{N+1}}{\lambda_{N}}-\frac{\lambda_{N}}{\lambda_{N-1}}\right).
\end{eqnarray*}
and thus the corresponding uncertainty relation is
\begin{eqnarray*}
\langle (\Delta X)^{2}\rangle\langle (\Delta Y)^{2}\rangle \geq
\left(\frac{\lambda_{N+1}}{\lambda_{N}}-\frac{\lambda_{N}}{\lambda_{N-1}}\right)^{2}.
\end{eqnarray*}
A state is said to be squeezed if either $\langle (\Delta X)^{2}\rangle$ or
$\langle (\Delta Y)^{2}\rangle$ is less than $\left(\frac{\lambda_{N+1}}{\lambda_{N}}-\frac{\lambda_{N}}{\lambda_{N-1}}\right)$.
To review the principle of squeezing, we consider an appropriate quadrature operator
\begin{eqnarray*}
X_{\theta} = X\cos\theta + Y\sin\theta = Ae^{-i\theta} +
A^{\dag}e^{i\theta}
\end{eqnarray*}
which gives
\begin{eqnarray*}
\begin{array}{rcl}
\langle(\Delta X_{\theta})^{2}\rangle
& = & (\langle  A^{2}\rangle - \langle A\rangle^{2})e^{-2i\theta}
+(\langle  {A^{\dag}}^{2}\rangle - \langle A^{\dag}\rangle^{2})e^{2i\theta}
+ \langle AA^{\dag}\rangle\\
&   & + \langle AA^{\dag}\rangle-\langle A\rangle\langle A^{\dag}\rangle
+\langle A^{\dag}A\rangle -\langle A^{\dag}\rangle\langle A\rangle
\end{array}
\end{eqnarray*}

After observing $\langle A\rangle = \overline{\langle A^{\dag}\rangle}$, we obtain
\begin{eqnarray*}
 \langle:(\Delta X_{\theta})^{2}:\rangle =
 \bar{\zeta}e^{-2i\theta} + \zeta e^{2i\theta} + 2\langle
A^{\dag}A\rangle - 2|\langle A^{\dag}\rangle|^{2}
\end{eqnarray*}
where $\zeta = \langle{A^{\dag}}^{2}\rangle - \langle
A^{\dag}\rangle^{2}$. Finally we have
\begin{eqnarray}
\begin{array}{rcl}
\label{eq15}
S_{\mathrm{opt}} & = & \langle:(\Delta X_{\theta})^{2}:\rangle_{\min}\\
& = & -2|\langle{A^{\dag}}^{2}\rangle - \langle
A^{\dag}\rangle^{2}|+2\langle A^{\dag}A\rangle - 2|\langle
A^{\dag}\rangle|^{2}
\end{array}
\end{eqnarray}

The expectations $\langle {A^{\dag}}^{2}\rangle$ and $\langle A^{\dag}\rangle$ with respect to the state vector $|\psi(t_1)\rangle$ (\ref{eq8}) can be calculated as
\begin{eqnarray}
\label{eq16}
\langle {A^{\dag}}^{2}\rangle
 =  2\sum_{n}\sqrt{\frac{\lambda_{n+1}}{\lambda_{n-1}}}C_{a,n-1}\bar{C}_{a,n-1}
+\sum_{n}\sqrt{\frac{\lambda_{n+2}}{\lambda_{n}}}C_{c,n}\bar{C}_{c,n}
\end{eqnarray}
and
\begin{equation}
\label{eq17}
\langle A^{\dag}\rangle
=  2\sum_{n}\sqrt{\frac{\lambda_{n}}{\lambda_{n-1}}}C_{a,n-1}\bar{C}_{a,n-1}
+ \sum_{n}\sqrt{\frac{\lambda_{n+1}}{\lambda_{n}}}C_{c,n}\bar{C}_{c,n}
\end{equation}

Substituting (\ref{eq16}) and (\ref{eq17}) in (\ref{eq15}), we have obtained an expression of $S_{\mathrm{opt}}$ for initial coherent state $F_{n}(0)=\exp(-\bar{n}/2)\frac{\bar{n}^{n/2}e^{i\zeta n}}{\sqrt{n!}}$. We have investigated the possibility of observing squeezing analytically and plotted $S_{\mathrm{opt}}$ as a function of scaled time $gt$ for $\bar{n}=0.3$, by choosing $\lambda_{n} \sim n!, (n!)^{2}$ and $[n]!$ with $[n]=(1-q^{n})/(1-q), 0<q<1$, respectively.

\begin{figure*}[ht]
\centering
\includegraphics[width=5cm]{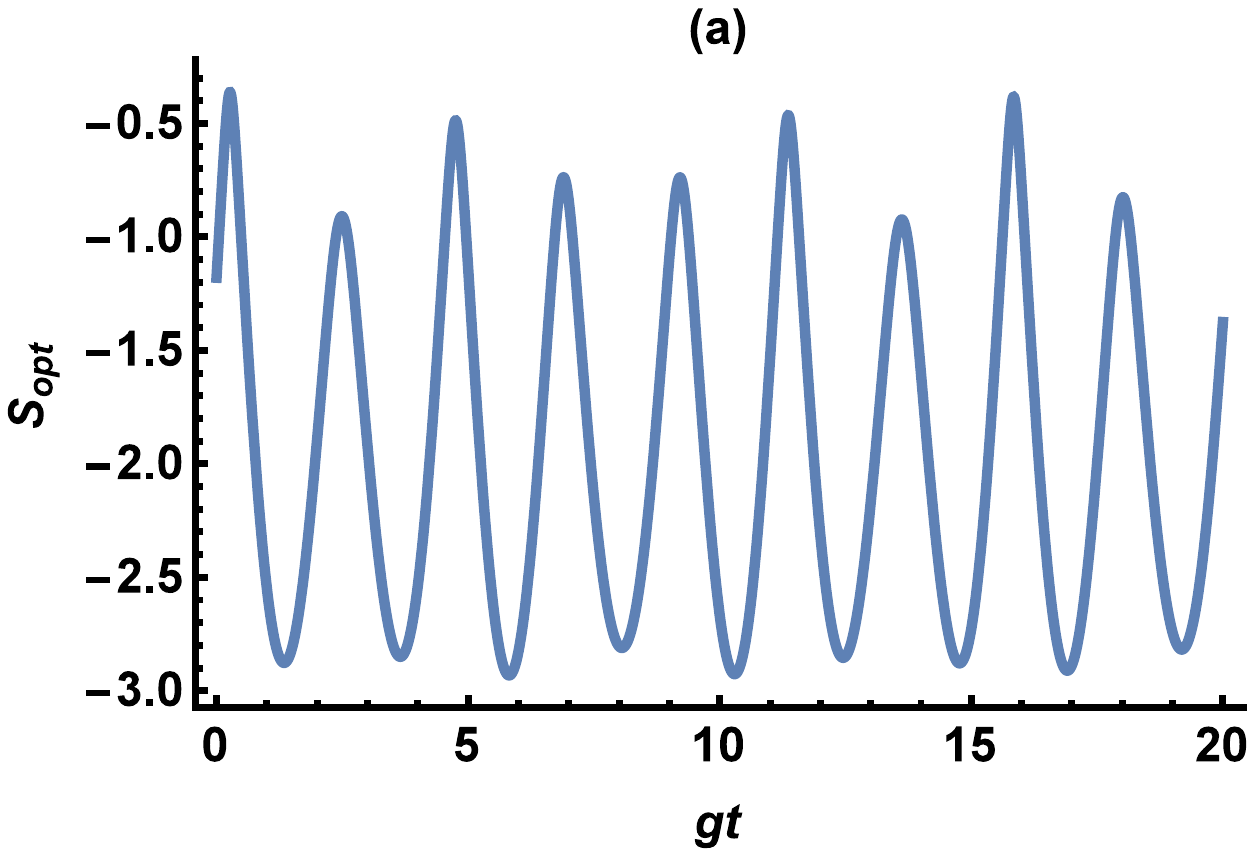}\hspace{1cm}
\includegraphics[width=5cm]{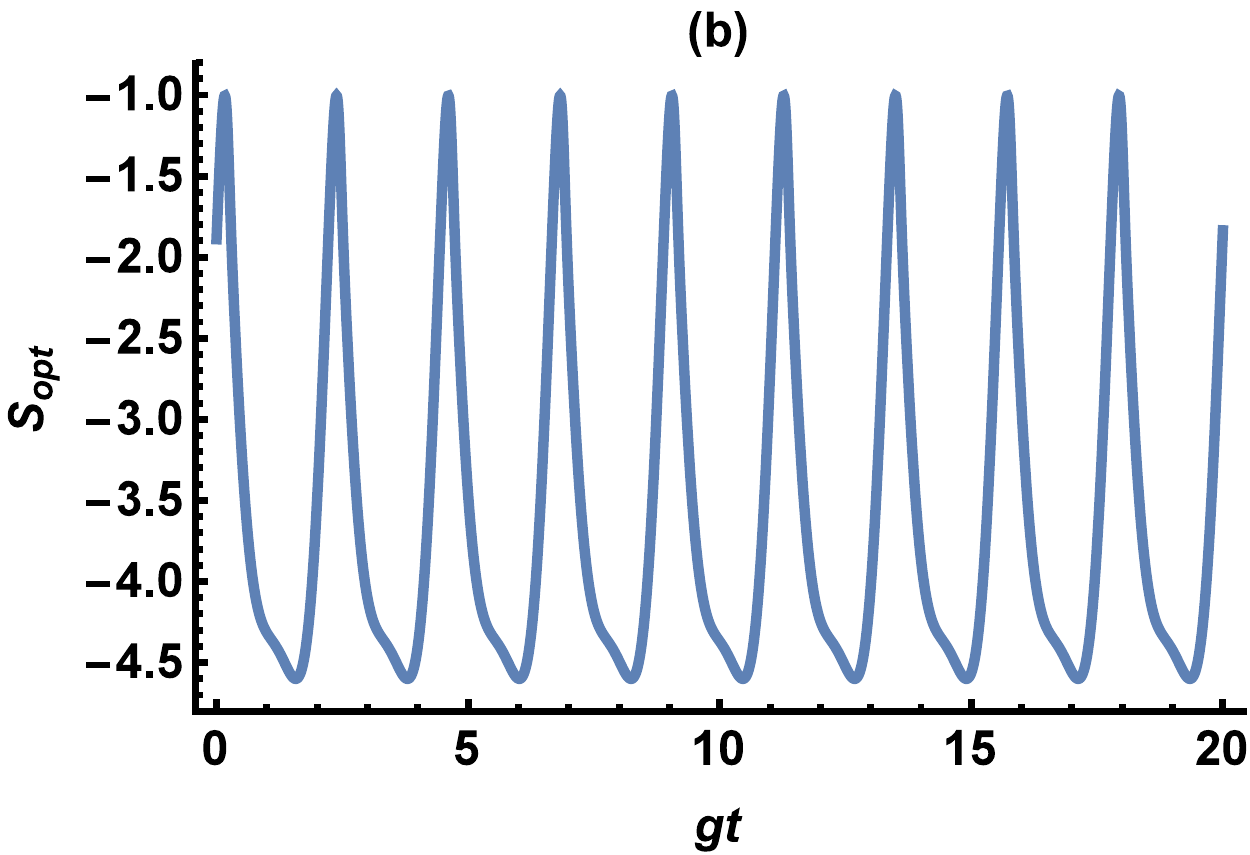}\hspace{1cm}
\includegraphics[width=5cm]{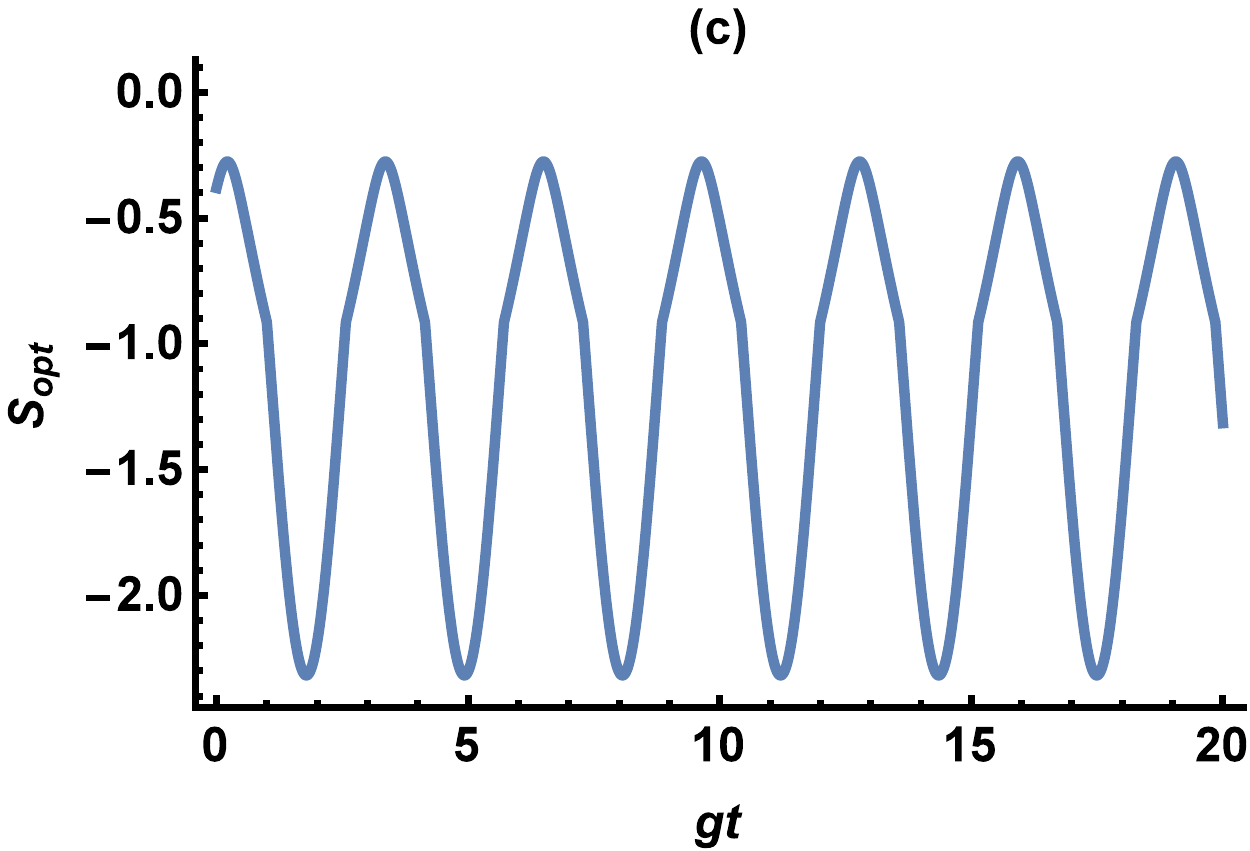}\hspace{1cm}

\caption{(Color online) Squeezing as a function of $gt$ for $\alpha_0=0.3$ and for different values of $\lambda_n$ such that $(a)~\lambda_n=n!$; $(b)~\lambda_n=(n!)^2$ and $(c)~\lambda_n=[n]$.}
\label{fig4}
\end{figure*}

\begin{figure*}[ht]
\centering
\includegraphics[width=5cm]{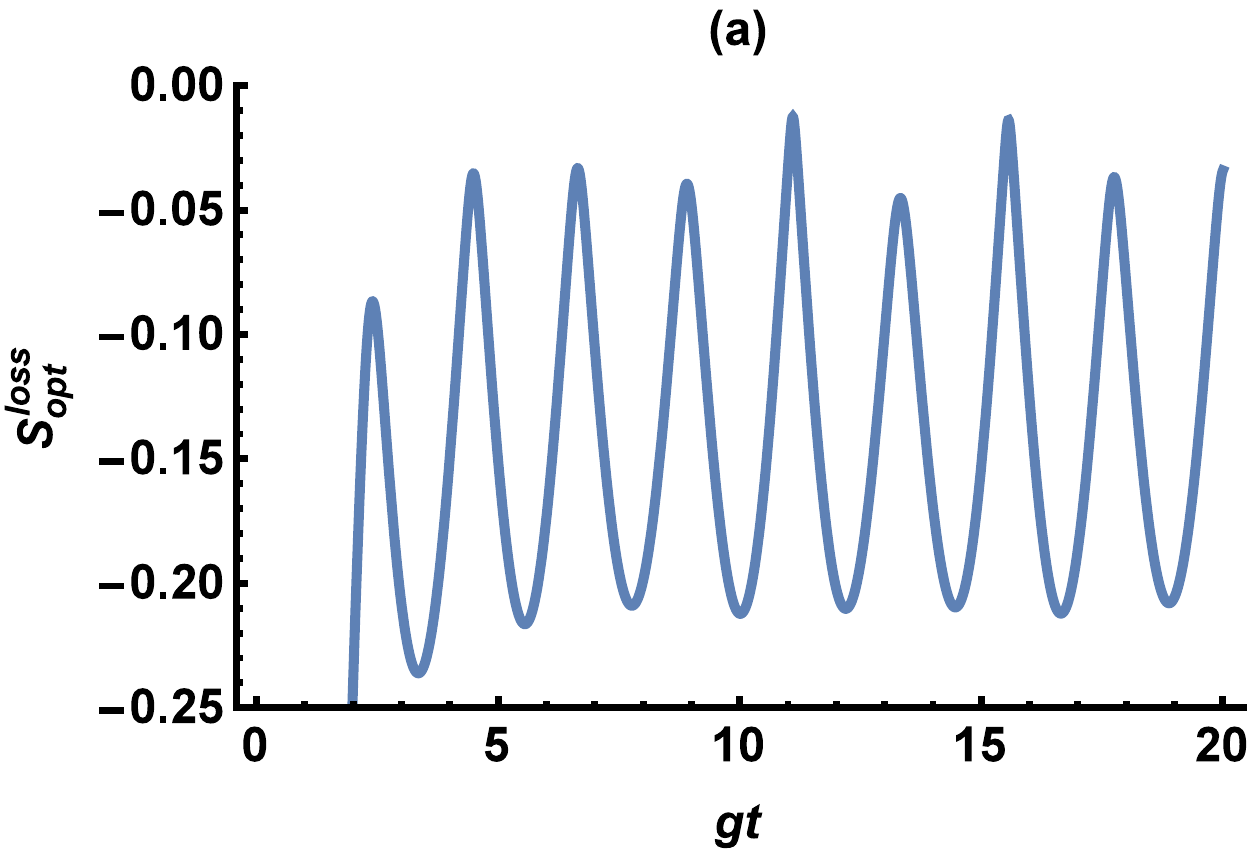}\hspace{1cm}
\includegraphics[width=5cm]{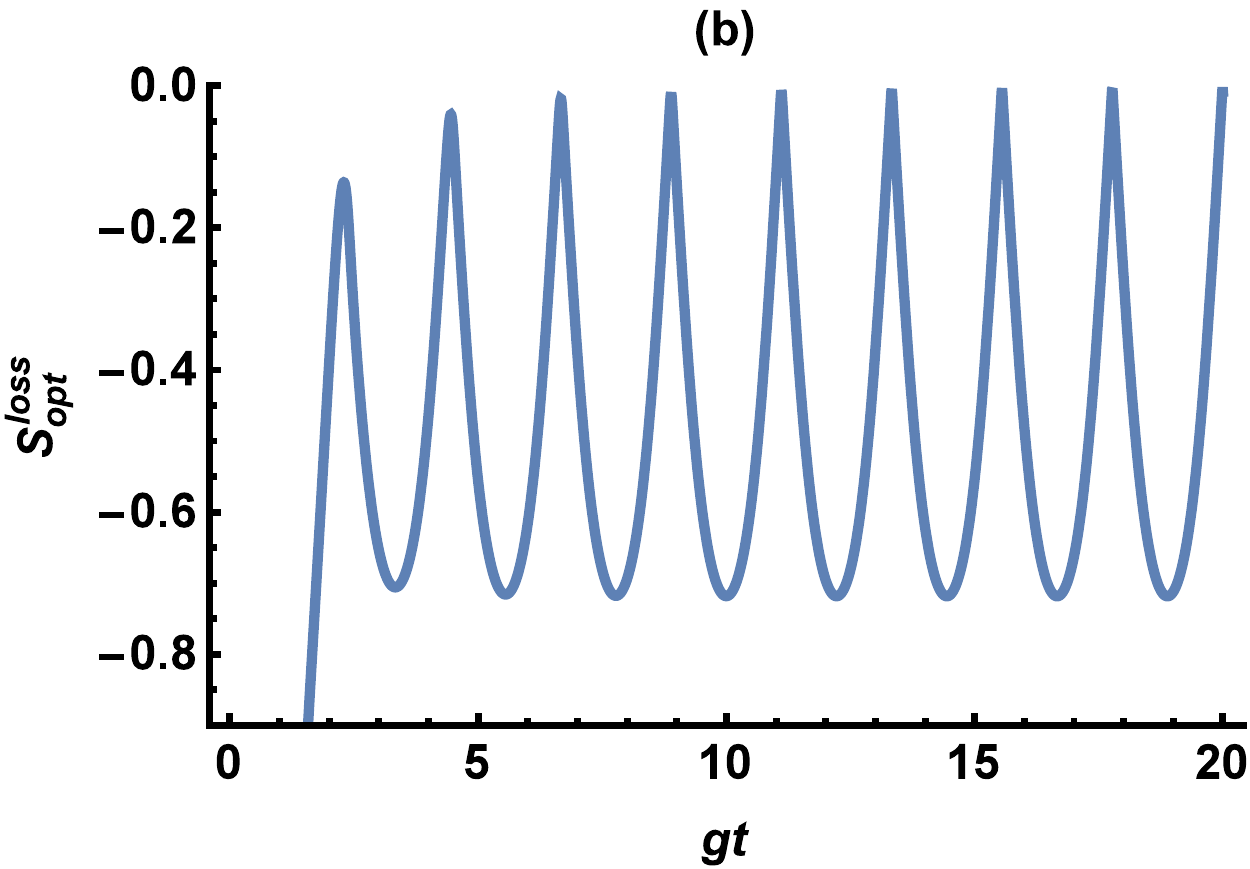}\hspace{1cm}
\includegraphics[width=5cm]{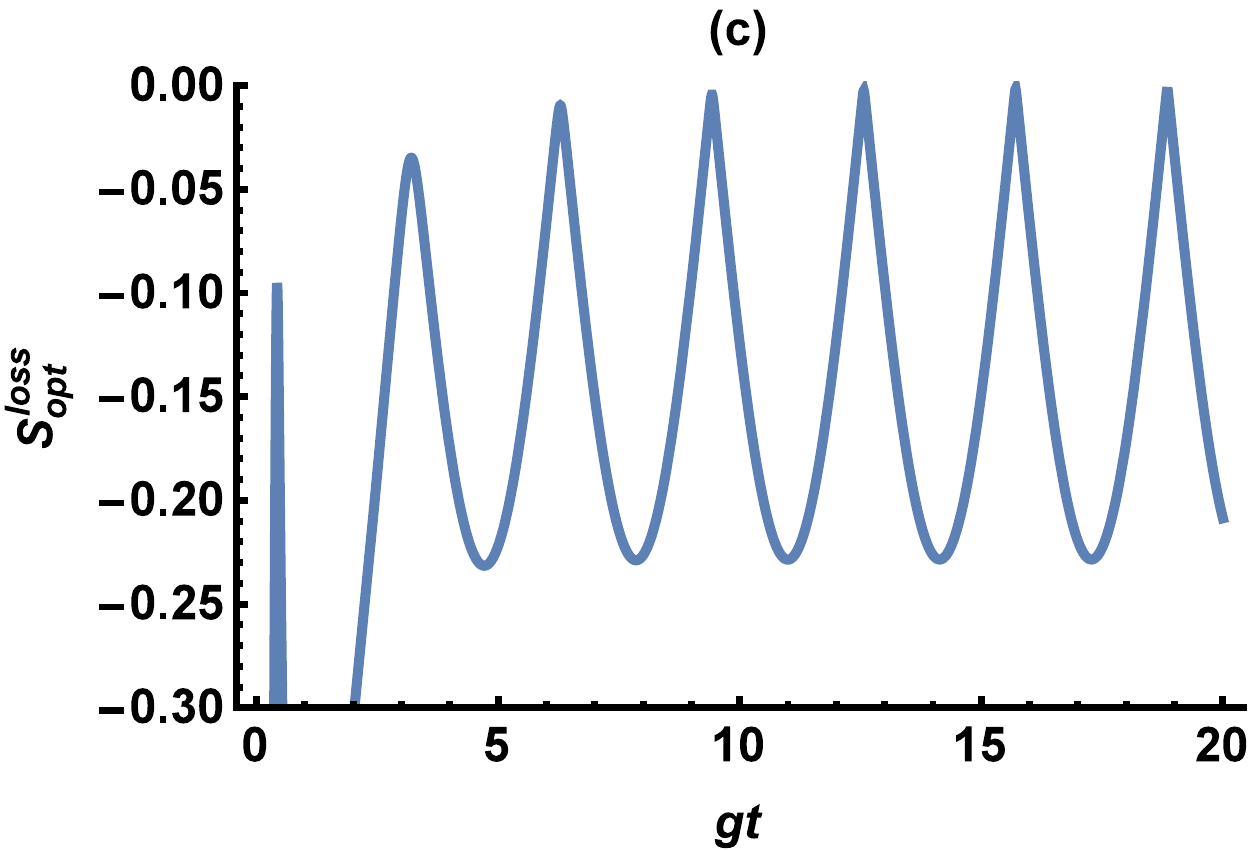}\hspace{1cm}

\caption{(Color online) Squeezing after loss as a function of $gt$ for $\bar{n}=0.3$, $k=0.5$ and for different values of $\lambda_n$ such that $(a)~\lambda_n=n!$; $(b)~\lambda_n=(n!)^2$ and $(c)~\lambda_n=[n]$.}
\label{fig5}
\end{figure*}

Fig.~\ref{fig4} describes the first-order squeezing in terms of the scaled time $gt$ for different $\lambda_n$. In all the cases, squeezing has been seen which is a clear evidence of the nonclassical nature of the radiation field. In Fig.~\ref{fig5}, we have illustrated squeezing parameter with cavity decay rate $k=0.5$. It is clear that the magnitude of squeezing lowers in presence of cavity decay. It is also observed that with this amount of decay, squeezing occurs while the negativity of $Q^M_{\mathrm{loss}}$ disappears [see Fig.~\ref{fig3}]. That means, the squeezing parameter is performing better than Mandel's $Q^M$ to detect the nonclassical character of the radiation field while loss of cavity arises.

\section{Conclusion}

In this paper, we have studied an atom-cavity field interaction in interacting Fock space. After giving a brief introduction of the interacting Fock space, we have found the state vector for the atom-field coupling. The radiation field becomes super-poissonian when $\lambda_n=(n!)^2$. The behavior of Mandel's $Q^M$ also reveals that the state remains almost nonclassical when $\lambda_n = [n]$. We have also plotted the normal squeezing against the scaled time $gt$ and shown that squeezing occurs for different $\lambda_n$ as $gt$ varies. We have also checked the effect of cavity decay rate over these nonclassicality witnesses. Our result establishes that cavity decay rate has an immense control over reducing the nonclassicality of the cavity field.  This work clearly demonstrates the dynamics of a three-level atom-cavity interaction in interacting Fock space and the effects of different parametric values of $\lambda_n$ over Mandel's $Q^M$ and normal squeezing.

The atom-cavity model is an extremely purposeful model from the perspective of quantum information theory (QIT) and quantum computation. The nonclassical interaction between the atomic levels and the cavity modes gives rise to entanglement, which is the primary resource for performing various QIT tasks such as superdense coding, quantum teleportation and quantum key distribution \cite{horo,bennett}. Another important aspect of studying this models is the distinct possibility of generating such bipartite interactions in the laboratory for investigating and applying the correlated system in future quantum tasks. Therefore understanding the basic dynamics of nonclassicality and related controlling factors in a simple generalized atom-field interaction is of utmost importance for realization of such quantum systems.

%


%
%

\bibliographystyle{spmpsci}      

\begin{thebibliography}{}

\bibitem{girish} G. S. Agarwal, Quantum Optics, Cambridge University Press, Cambridge, (2013).
\bibitem{das5} P. K. Das and P. Haldar, Infinite Dimensional Analysis, Quantum Probability and Related Topics, Vol. 16 (2), 1769-1778 (2013).

\bibitem{peng1} J. S. Peng, G. X. Li and P. Zhou, Phys. Rev. A, Vol. 46 (3), 1516 (1992).
\bibitem{ekert}  D. Bouwmeester, A. Ekert and A. Zeilinger, The Physics of Quantum Information,
Springer, Berlin, 2000.

\bibitem{hillery00} M. Hillery, Phys. Rev. A, Vol. 61(2), 022309 (2000).

\bibitem{akira98} A. Furusawa, J. L. S{\o}rensen, S. L. Braunstein, C. A. Fuchs, H. J. Kimble and E. S. Polzik, Science, Vol. 282 (5389), 706–709 (1998).

\bibitem{abbott16} B. P. Abbott, R. Abbott et. al., Phys. Rev. Lett., Vol. 116(6), 061102 (2016).

\bibitem{harry10} G. M. Harry, Classical and Quantum Gravity, Vol. 27(8), 084006 (2010).
\bibitem{grote13} H. Grote, K. Danzmann, K. L. Dooley, R. Schnabel, J. Slutsky and H. Vahlbruch, Phys. Rev. Lett., Vol. 110(18), 181101 (2013).
\bibitem{zeilinger} A. Zeilinger, Rev. Mod. Phys., Vol. 71, S288 (1998).
\bibitem{kishore14} K. Thapliyal, A. Pathak, B. Sen and J. Pe\v{r}ina, Phys. Rev. A, Vol. 90(1), 013808 (2014).
\bibitem{kishore17} S. Giri, K. Thapliyal, B. Sen and A. Pathak, Physica A: Stat. Mech. Appl., Vol. 466, 140–152 (2017).
\bibitem{nasir16} N. Alam and S. Mandal, Opt. Commun., Vol. 359, 221–233 (2016).
\bibitem{loudon} R. Loudon, Phys. Bull., Vol. 27 (1), 21 (1976).
\bibitem{malvin} M. C. Teich and B. E. A. Saleh, Prog. Opt., Vol. 26, 1-104 (1988).

\bibitem{ghatak} A. Pathak and A. Ghatak, JEWA, Vol. 32 (2), 229-264 (2018).
\bibitem{nasirn1} N. Alam, A. Verma and A. Pathak, Phys. Lett. A, Vol. 382 (28), 1842-1851 (2018).
\bibitem{lounis} B. Lounis and M. Orrit, Rep. Prog. Phys., Vol. 68 (5), 1129-1180 (2005).
\bibitem{nasirn2} P. Malpani et. al., Opt. Commun., Vol. 459, 124964 (2020).
\bibitem{nasirn3} P. Malpani et. al., Annalen der Physik, Vol. 532 (1), 1900337 (2020).

\bibitem{nasirn4} P. Malpani et. al., Annalen der Physik, Vol. 531 (11), 1900141 (2019).
\bibitem{lee1} C. T. Lee, Phys. Rev. A, Vol. 42 (3), 1608 (1990).



\bibitem{jay} E. T. Jaynes and F. W. Cummings, Proc. IEEE, Vol. 51, 89 (1973).

\bibitem{sten} S. Stenholm, Phys. Rep., Vol. 6, 1 (1973).
\bibitem{eberly} J. H. Eberly, N. B. Narozhny and J. J. Sanchez-Mondragon, Phys. Rev. Lett., Vol. 44, 1323 (1980).
\bibitem{martin} M. Bouillard et. al., Opt. Exp., Vol. 27 (3), 3113 (2019).



\bibitem{arpita19} A. Chatterjee, J. Mod. Opt., Vol. 66 (8), 898-908 (2019).
\bibitem{accar1} L. Accardi L. and Y. G. Lu Y.G., Commun. Math. Phys., Vol. 180, 605-632 (1996).
\bibitem{accar2} L. Accardi, Y. G. Ku and I. Volvovich, Quantum theory and its Stochastic Limit, Springer,2001.

\bibitem{nasir1} N. Alam, K. Mandal and A. Pathak, Int. J. Theo. Phys., Vol. 57, 3443–3456 (2018).


\bibitem{fle05} M. Fleischhauer, A. Imamoglu and J. P. Marangos, Rev. Modern Phys., Vol. 77, 633 (2005).
\bibitem{scully1} M. O. Scully, S. -Y. Zhu and A. Gavrielides, Phys. Rev. Lett., Vol. 62, 2813 (1989).
\bibitem{scully2} M. O. Scully and M. S. Zubairy, Quantum Optics, Cambridge University Press, Cambridge, (1997).
\bibitem{sandhya} S. N. Sandhya and V. Ravishankar, Phys. Rev. A, Vol. 82, 062301 (2010).
\bibitem{arpita12} A. Chatterjee, Phys. Lett. A, Vol. 376, 1601 (2012).

\bibitem{luigi1} L. Accardi and M. Bozejko, Inf. Dimn. Analy., QP and related topics, Vol. 1 (4), 663-670 (1998).

\bibitem{luigi2} L. Accardi and P. K. Das, Int. J. Theo. Phys., Vol.42 (11), 2721-2734 (2003).

\bibitem{luigi3} L. Accardi, V. Crismale and Y. G. Lu, Inf. Dimn. Analy., QP and related topics, Vol.8 (4), 631-650 (2005).

\bibitem{luigi4} L. Accardi, H. H. Kuo and A. Stan, Inf. Dimn. Analy., QP and related topics, Vol. 10 (4), 591-612 (2007).

\bibitem{das1} P. K. Das,  QP-PQ: Quantum Probability and White Noise Analysis, Vol. 18, 141-152 (2005).

\bibitem{das2} P. K. Das, Int. J. Theo. Phys., Vol. 41 (6), 1099-1106 (2002).

\bibitem{das3} P. K. Das, Int. J. Theo. Phys., Vol. 41 (10), 2013-2024 (2002).

\bibitem{das4} P. K. Das and P. Haldar, Mod. Phys. Lett. B, Vol. 25 (21), 1769-1778 (2011).

\bibitem{peng2} J. S. Peng and G. X. Li, Phys. Rev. A, Vol. 45 (5), 3289 (1992).

\bibitem{mandel} L. Mandel, Opt. Lett., Vol. 4 (7), 205-207 (1979).

\bibitem{cohen} C. Cohen‐Tannoudji, J. Dupont‐Roc and G. Grynberg, Atom—Photon Interactions: Basic Process and Appilcations, Wiley-VCH Verlag, (1998).

\bibitem{arpita3} A. Ghosh and P. K. Das, Can. J. Phys., Vol. 86, 811-818, (2008).
\bibitem{sb} S. B. Li, X. B. Zou and G. C. Guo. Phys. Rev. A, Vol. 75, 045801 (2007).

\bibitem{deng} Z. J. Deng, M. Feng and K. L. Gao, Phys. Rev. A, Vol. 73, 014302 (2006).
\bibitem{arpita2} A. Ghosh and P. K. Das, Int. J. Theo. Phys., Vol. 47, 1731-1741 (2008).

\bibitem{walls} D. F. Walls and G. J. Milburn, Quantum Optics, Springer-Verlag, Berlin (1994).
\bibitem{lee2} S. Y. Lee and H. Nha, Phys. Rev. A, Vol. 82, 053812 (2010).

\bibitem{wang1} X. Wang and Barry C. Sanders, Phys. Rev. A, Vol. 68, 033821 (2003).
\bibitem{horo} R. Horodecki, P. Horodecki, M. Horodecki and K. Horodecki, Rev. Mod. Phys., Vol. 81, 865 (2009).
\bibitem{bennett} C. H. Bennett and S. J. Wiesner, Phys. Rev. Lett., Vol. 69, 2881 (1993).
\end{thebibliography}


\end{document}